\documentstyle[aps,pre,epsfig]{revtex}

\newcommand{\ba}{\begin{eqnarray}}
\newcommand{\be}{\begin{equation}}
\newcommand{\ea}{\end{eqnarray}}
\newcommand{\ee}{\end{equation}}

\begin{document}

\draft

\twocolumn[\hsize\textwidth\columnwidth\hsize\csname@twocolumnfalse%
\endcsname

\title{Liquid antiferromagnets in two dimensions}
\author{Carsten Timm\cite{email}}
\address{Institut f\"ur Theoretische Physik, Freie Universit\"at
Berlin, Arnimallee 14, D-14195 Berlin, Germany}
\date{February 22, 2002}

\maketitle

\begin{abstract}
It is shown that, for proper symmetry of the parent lattice,
antiferromagnetic order can survive in two-dimensional liquid crystals and
even isotropic liquids of point-like particles, in contradiction to what
common sense might suggest. We discuss the requirements for
antiferromagnetic order in the absence of translational and/or
orientational lattice order. One example is the honeycomb lattice, which
upon melting can form a liquid crystal with quasi-long-range orientational
and antiferromagnetic order but short-range translational order. The
critical properties of such systems are discussed. Finally, we draw
conjectures for the three-dimensional case.
\end{abstract}


]

\narrowtext


Ferrofluids, {\it i.e.}, suspensions of small ferromagnetic particles in a
carrier liquid, have been studied quite extensively \cite{ferrofl}. These
materials are really liquid superparamagnets without long-range magnetic
order in the absence of an applied magnetic field. However, there is no
fundamental reason why true ferromagnetism should not exist in a liquid. The
strong, short-range exchange interactions are not strongly affected by the
absence of crystalline order, as shown by the existence of amorphous
ferromagnets \cite{amof}.

The present paper addresses the question of whether {\it antiferromagnetic
liquids}, which one could call ``antiferrofluids,'' are also possible. On
first sight, the answer seems to be no. Common sense tells us that the huge
frustration in a liquid destroys antiferromagnetic order. To construct an
antiferromagnetic liquid one would thus look for liquids that partially
retain structural order, {\it i.e.}, liquid crystals. In fact,
antiferro{\it electric\/} liquid crystals have been studied extensively
\cite{AFE.ex}. These materials consist of long, polar molecules so that
antiferroelectric order appears rather naturally in their smectic phases.

The question we want to discuss here is whether liquids (including liquid
crystals) consisting of {\it spherical\/} particles with a spin degree of
freedom can sustain antiferromagnetic order. At least in two dimensions this
is possible, as we show below. We consider two-dimensional (2D) systems,
since in two dimensions the theory of melting is much further developed than
in three. The relevance for three-dimensional systems is briefly discussed
afterwards. We introduce spin anisotropy to obtain a finite-temperature
phase transition. Specifically, we think of the antiferromagnetic order
parameter having either {\it XY\/} or Ising symmetry. In the first case
there is a Berezinkii-Kosterlitz-Thouless (BKT) transition \cite{BKT} and
the low-temperature phase has quasi-long-range order. In the second case
there is an Ising-type transition \cite{Ising} to a long-range-ordered
phase.


Our arguments employ the theory of 2D melting developed by Nelson, Halperin,
and Young \cite{NHY} (NHY), which is based on the BKT renormalization group
theory \cite{BKT,Minn}. We first briefly review this theory. Then we discuss
melting of a lattice with antiferromagnetic order for the normal case that
antiferromagnetism is strongly frustrated by melting \cite{TGF}. This sets
the stage for the discussion of the possibility of antiferromagnetism in the
liquid crystal formed upon melting. Surprisingly, for certain lattices
melting can even produce an {\it isotropic\/} liquid that retains
antiferromagnetic order.


The NHY theory \cite{NHY,CL} predicts two distinct melting transitions. The
one at the lower temperature separates a 2D solid with quasi-long-range
translational order from a liquid crystal with short-range translational but
quasi-long-range orientational order \cite{rem.spec}. This transition is due
to the unbinding of pairs of dislocations. Dislocations are point-like in 2D
and can be thermally created in pairs or multiplets of vanishing total
Burgers vector. Pairs of dislocations with opposite Burgers vector have an
attractive logarithmic interaction, similar to vortex-antivortex pairs in
the 2D {\it XY\/} model. The resulting BKT-type transition is characterized
by a jump of Young's modulus (the stiffness against tension), which is
finite and universal just below the transition and zero above. In the
liquid-crystal phase bound pairs of {\it disclinations\/} exist, which are
defects of the orientational order. This order is destroyed at a higher
transition temperature where disclination pairs unbind. Since their
interaction is logarithmic in the presence of free dislocations, the
transition is also of BKT-type. Note that one or both transitions {\it
may\/} be replaced by a first-order transition.

\begin{figure}[h]
\epsfig{file=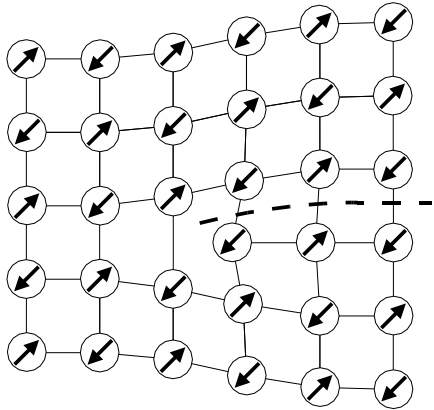,width=3in}
\caption{Square lattice antiferromagnet with an elementary dislocation. The
magnetic order is maximally frustrated along the heavy dashed line. For {\it
XY\/} spins this configuration is unfavorable and the spins will relax to
spread out the frustration \protect\cite{TGF}. As the result, the
dislocation dresses with half a vortex.}
\label{fig.sq1}
\end{figure}

What happens if the particles carry a spin with a tendency to order
antiferromagnetically? We restrict ourselves to bipartite lattices. Then
the spins show N\'eel order in the classical ground state, if frustrating
longer-range interactions are not too strong. For most simple lattices such
as the square lattice elementary dislocations \cite{rem.elem} frustrate the
magnetic order, as illustrated by Fig.~\ref{fig.sq1}. There is a line of
maximally frustrated bonds ending at the dislocation. This line could end
at another dislocation of opposite Burgers vector. The energy of such a
pair is {\it linear\/} in their separation and the pair is
confined. This is indeed the case for Ising spins \cite{Cardy}. On the
other hand, for two-component ({\it XY\/}) spins Fig.~\ref{fig.sq1} does not
show the lowest-energy configuration. Rather, the spins relax to spread the
frustration more evenly. In effect, the dislocation dresses with {\it half a
vortex\/} (or antivortex) in the N\'eel order \cite{TGF}. The dislocation
interaction is now again logarithmic, but with a contribution from the
half vortices. The interplay of dislocation-unbinding and magnetic
transitions in this case has been studied in Ref.~\cite{TGF}. It is obvious
that magnetic order cannot survive the dislocation unbinding, since free
dislocations carry (fractional) vorticity and act like free vortices
\cite{BKT,TGF}. Of course, the magnetic transition {\it may\/} take place at
a {\it lower\/} temperature than the dislocation unbinding.

\begin{figure}[h]
\epsfig{file=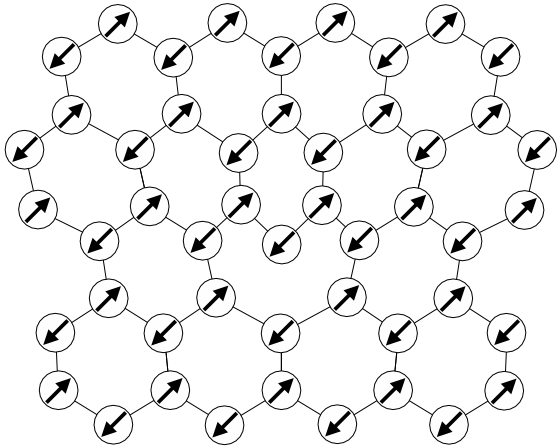,width=3in}
\caption{Honeycomb lattice antiferromagnet with an elementary dislocation.
Evidently the defect does not frustrate the magnetic order and,
consequently, does not dress with fractional vorticity.}
\label{fig.hc1}
\end{figure}

However, antiferromagnetism need not be destroyed at the lower melting
temperature, if dislocations do not frustrate the magnetic order. One
example is the honeycomb lattice. An elementary dislocation \cite{rem.elem}
does not frustrate the antiferromagnet, as shown in Fig.~\ref{fig.hc1}.
Since all possible dislocations are superposition of elementary ones, none
of them frustrates the order. Consequently, free dislocations above the
lower melting temperature do not carry vorticity and thus the existence of
free dislocations does not preclude antiferromagnetic (long-range or
quasi-long-range) order \cite{rem.honeymelt}.

When do dislocations not frustrate the magnetic order? This is the case if
their Burgers vectors connect two sites with the same spin direction, {\it
i.e.}, on the same sublattice. The Burgers vector can be any lattice vector
of the lattice without spins. Hence, all dislocations do not frustrate if
any translation by a lattice vector leaves the spins invariant. Or, in other
words, if magnetic ordering does not reduce the set of translational
symmetry operations of the lattice. This is the case for the honeycomb
lattice, which already has a two-site basis. On the other hand, for the
square lattice the order reduces the set of translations and
dislocations exist that frustrate the magnetic order.

We now turn to the upper, disclination-unbinding transition. For the
honeycomb lattice, disclinations are characterized
by the angle modulo $2\pi$ by which the bond angle changes if one goes
around the defect \cite{Klei,CL}.
The elementary disclinations \cite{rem.elem} of the honeycomb lattice and
the corresponding liquid crystal are $\pm2\pi/6$ disclinations centered at
a hexagonal plaquette. Thus, the defects have a five- or seven-sided
plaquette at their core, which obviously frustrates the magnetic order.
Furthermore, there are paths of arbitrarily large length around the defect
that consist of an odd number of bonds. For the {\it XY\/} model, the spins
again relax to reduce the energy and the disclinations dress with half
vortices. Consequently, the magnetic transition temperature cannot lie above
the disclination-unbinding temperature.

\begin{figure}[h]
\epsfig{file=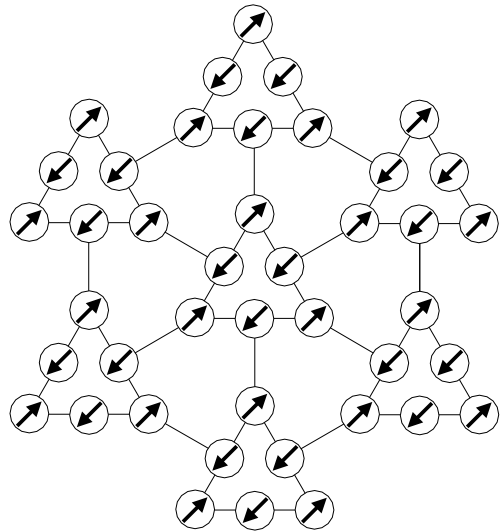,width=3in}
\caption{A more complicated lattice. Here neither
dislocations nor disclinations frustrate the magnetic order.}
\label{fig.tr0}
\end{figure}
\begin{figure}[h]
\epsfig{file=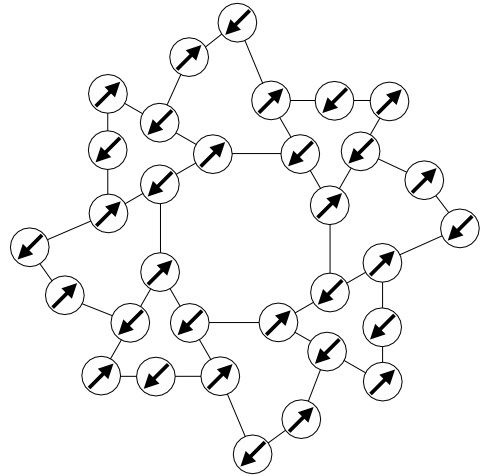,width=3in}
\caption{The core of a $+2\pi/3$ disclination for the lattice shown in
Fig.~\protect\ref{fig.tr0}. The defect does not frustrate the magnetic
order.}
\label{fig.tr2}
\end{figure}

The next question is whether there are lattices for which neither
dislocations nor disclinations frustrate the magnetic order. The lattice in
Fig.~\ref{fig.tr0} satisfies the criterion for non-frustrating
dislocations. Furthermore, elementary disclinations with a change of the
bond angle by $\pm2\pi/3$ do not frustrate either, as illustrated by
Fig.~\ref{fig.tr2}. If the appearance of magnetic order does not reduce the
{\it orientational\/} symmetry, {\it i.e.}, does not remove rotation axes
or reduce their multiplicity, all disclinations are compatible with
antiferromagnetic order. In this case antiferromagnetic order {\it can\/}
exist in the isotropic liquid above the upper melting transition. There is
another way to express the condition for the existence of non-frustrating
dislocations and disclinations for bipartite lattices: Magnetic order in
the isotropic liquid is possible if the corresponding lattice does have
{\it two non-equivalent sublattices}, {\it i.e.}, one cannot be mapped onto
the other by any translation or rotation or combination thereof. Then
antiferromagnetic ordering does not reduce the lattice symmetry. At higher
temperatures the liquid should eventually loose the hidden order that is
expressed by the non-equivalence of two subsystems. Note, however, that
this cannot happen through disclination unbinding, but will probably
take place at a first-order transition.

Even if dislocations (or disclinations) do not dress with vorticity, their
energies depend on the magnetic order, since part of the interaction is of
magnetic origin. Conversely, due to frustration of the magnetic interaction
at larger distances structural order affects the vortex energies. We now
argue why this subdominant coupling leaves the principal picture unchanged,
focusing on dislocations and vortices. The interaction energy of
dislocations is proportional to Young's modulus, which we expect to be a
continuous function of the vortex density. Since the vortex density itself
is a continuous function of temperature through the vortex-unbinding
transition \cite{BKT,CThigh}, the parameters entering in the BKT theory of
dislocation unbinding are continuous through the magnetic transition. A
similar argument can be made for the change of the vortex energy due to
dislocations. If one tunes the strength of magnetic vs.\ non-magnetic
interactions, the dislocation-unbinding and vortex-unbinding transitions
thus cross in a tetracritical point and both the structural and the magnetic
order show a universal BKT jump at this point. This is drastically different
from the normal case of, {\it e.g.}, the square lattice, where for strong
magnetic interactions the two transitions merge into a single one, at which
only one order parameters shows a {\it universal\/} jump \cite{TGF}.

Next, we briefly commend on the low-energy collective excitations of
liquids with antiferromagnetic quasi-long-range order. First, there is the
usual longitudinal acoustic phonon branch.
The liquid crystal phases differ from the isotropic liquid in that they have
massive topological excitations, {\it i.e.}, the disclinations. In addition,
for {\it XY\/} spins there is a linearly dispersing spin wave mode in all
liquid phases. Its presence is one main characteristic of an
antiferromagnetic liquid. This mode leads to the characteristic behavior
of the magnetic susceptibility of an antiferromagnet \cite{AM}.

How can these considerations be applied to three-dimensional systems? In
three dimensions melting typically proceeds by a first-order transition
directly to the isotropic liquid. Nevertheless a dislocation-unbinding
mechanism may apply \cite{Mott,3Drev,BPS,Klei}. To obtain an isotropic
liquid, disclinations also have to unbind \cite{Klei}. They usually do so
at the same temperature, but this does not invalidate our criterion for
antiferromagnetic fluids. Note also that our arguments never took advantage
of the two-dimensionality. Thus it may be inferred that also in three
dimensions antiferromagnetic liquids can exist if the underlying lattice
has {\it two inequivalent sub-lattices}.


Finally, we turn to possible experimental realizations. A soft 2D {\it
XY\/} antiferromagnet is the Skyrmion crystal in the quantum Hall system
close to filling factor $\nu=1$ \cite{Letc,QHground,Slattice,TGF}.
Skyrmions are topological excitations of the ferromagnetic quantum Hall
state, which carry a quantized electric charge. Upon changing the filling
factor away from $\nu=1$, the extra charge appears in the form of
Skyrmions. The in-plane magnetization of a Skyrmion has a vortex-like
structure. Its direction can be characterized by a single {\it XY\/} angle
$\theta$, which couples antiferromagnetically \cite{Sachdev,Cote97}. The
classical ground states of the Skyrmion system are various lattice types
\cite{TGF,rem.qmelt}. One is a honeycomb lattice, albeit probably outside
of the realistic parameter range.

A straightforward realization of an Ising pseudospin model is a binary
alloy (in 2D or 3D). Another example is a system of vortices and
antivortices, which are prevented from annihilating, {\it e.g.}, by an
additional Coulomb repulsion. The vorticity then constitutes the Ising
degree of freedom. It has been suggested that such a vortex system is
formed when holes are doped into the antiferromagnetic cuprates
\cite{Vetc,TB}. These charged vortices might form a strongly anisotropic
(stripe) crystal at low temperatures \cite{WS,EK,TB}. It should be
interesting to apply the ideas of the present paper to its melting
\cite{EK}.


To conclude, we have shown that there is no fundamental reason why 2D, and
possibly 3D, antiferromagnetic liquids should {\it not\/} exist. Their
existence is determined by the structure of the underlying lattice: If
dislocations do not frustrate the antiferromagnetic order, antiferromagnetic
liquid-crystal phases are possible. One example is the honeycomb lattice.
If, in addition, disclinations also do not frustrate the magnetic order, it
can even survive in isotropic liquids. The crystal phase must have two
inequivalent sublattices for this to be possible. The resulting
``antiferrofluids'' would support spin waves with linear dispersion besides
longitudinal phonons. It would be worthwhile to search for experimental
realizations of this new phase of matter.


\end{document}